\documentclass[aps, twocolumn,amsmath,amstex,amssymb,mathfonts,superscriptaddress,longbibliography]{revtex4-2}

\usepackage{subfigure}
\usepackage{graphicx}
\usepackage{dcolumn}

\usepackage{physics}
\usepackage{bm}
\usepackage[colorlinks, linkcolor=cyan, anchorcolor=blue, citecolor=green]{hyperref} 
\usepackage{xcolor}

\usepackage{color}

\usepackage{amsthm}
\usepackage{amsfonts}
\usepackage{multirow}
\usepackage{makecell}
\usepackage{float}
\usepackage{comment}
\usepackage{enumitem}
\usepackage{verbatim}
\usepackage{mathtools}

\begin{document} 

\title{Melting of quantum Hall Wigner and bubble crystals}

\author{Haoyu Xia}
\thanks{These authors contribute equally to the work.}
\affiliation{International Center for Quantum Materials, Peking University, Beijing 100871, China}

\author{Qianhui Xu}
\thanks{These authors contribute equally to the work.}
\affiliation{Division of Physics and Applied Physics, Nanyang Technological University, Singapore 637371}

\author{Jiasen Niu}
\affiliation{Leiden Institute of Physics, Leiden University, 2333 CA Leiden, The Netherlands}

\author{Jian Sun}
\affiliation{International Center for Quantum Materials, Peking University, Beijing 100871, China}

\author{Yang Liu}
\affiliation{International Center for Quantum Materials, Peking University, Beijing 100871, China}

\author{L. N. Pfeiffer}
\author{K. W. West}
\affiliation{Department of Electrical Engineering, Princeton University, Princeton, New Jersey 08544, USA}

\author{Pengjie Wang}
\affiliation{Department of Physics, University of Illinois Urbana-Champaign, Urbana, Illinois 61801, USA}
\affiliation{Materials Research Laboratory, University of Illinois Urbana-Champaign, Urbana, Illinois 61801, USA}

\author{Bo Yang}
\email{yang.bo@ntu.edu.sg}
\affiliation{Division of Physics and Applied Physics, Nanyang Technological University, Singapore 637371}

\author{Xi Lin}
\email{xilin@pku.edu.cn}
\affiliation{International Center for Quantum Materials, Peking University, Beijing 100871, China}
\affiliation{Hefei National Laboratory, Hefei 230088, China}
\affiliation{Interdisciplinary Institute of Light-Element Quantum Materials and Research Center for Light-Element Advanced Materials, Peking University, Beijing 100871, China}

\begin{abstract}
A two-dimensional crystal melts via the proliferation and unbinding of topological defects, yet quantitatively predicting the melting temperature $T_m$ in real systems is challenging. Here we resolve this discrepancy in quantum Hall electron bubble phases by combining Corbino-geometry transport experiment in an ultraclean GaAs/AlGaAs quantum well for Landau levels $2\sim5$ with Hartree–Fock elasticity and the full Kosterlitz--Thouless--Halperin--Nelson--Young (KTHNY) melting criterion including the finite-temperature renormalization-group calculation. The theoretically obtained $T_m$ quantitatively captures the measured solid-liquid phase transition boundaries across all probed ranges, validating the bubble-crystal interpretation and establishing defect-mediated melting as a predictive framework for strongly interacting electronic solids. This agreement further supports using bulk transport to probe the energetics of topological defects and screening in quantum Hall physics, and the approach is readily extendable to other electronic crystals, including the generalized Wigner crystal in moiré Chern bands.
\end{abstract}

\maketitle 

\textit{Introduction--}
Although melting is a ubiquitous phenomenon, the melting points of only a small portion of solids have been accurately measured due to rigid experimental requirements, and reliable theoretical predictions are often nontrivial. The challenge becomes particularly acute at ultralow temperatures, where melting is increasingly dominated by quantum fluctuations rather than thermal phonons. In two dimensions (2D), the situation is further constrained fundamentally; there, true long-range crystalline order does not exist at any finite temperature due to the Mermin-Wagner theorem~\cite {PhysRev.176.250}. Kosterlitz-Thouless-Halperin-Nelson-Young (KTHNY) theory establishes that 2D melting is a continuous phase transition driven by the proliferation and unbinding of topological defects~\cite{Kosterlitz1973, halperin1978theory, nelson1979dislocation, young1979melting, Fisher1979prb}. Its quantitative predictions have been verified in classical Wigner crystals (WC), where the melting temperature $T_m$ exhibits the universal KTHNY scaling $T_m \propto \sqrt{n}$ ($n$ is the electron's density) with electron-electron interaction being Coulomb~\cite{PhysRevLett.42.795, Thouless1978, fisher1979phonon, morf1979temperature}. However, Wigner solids in strong magnetic fields at ultralow temperatures violate this scaling and display a nontrivial density dependence of $T_m$~\cite{Chen2006, Deng2012, Deng2012a, shingla2021particle}. In these strongly correlated systems, the lattice is largely softened by the strong quantum and thermal fluctuations, resulting in the overestimation of stiffness and $T_m$. Recent technical advances in van der Waals materials have facilitated the direct imaging of 2D electron solids and revealed the intertwined fluctuations and phase competition in the system, indicated by the short-range crystallinity and pronounced lattice softening~\cite{Li2024NatNano_SSEC, Li2021Nature_GWCimaging, Li2021, Tsui2024}. These observations underscore the complexity of quantum electronic lattices. 

The 2D electron bubble phases in ultra-high-mobility GaAs heterostructures provide an ideal platform for investigating these issues. Like the conventional Wigner crystal, a bubble phase is an interaction-driven electron solid, but features a generalized lattice where multiple electrons could cluster within one unit cell, which is predicted to be the ground state of a two dimensional electron gas (2DEG) under a perpendicular magnetic field at high Landau levels (LLs)~\cite{Moessner1996, shibata2001ground, Lewis2004, Deng2012, Deng2012a, rossokhaty2016electron, gabor2019, gabor2020, Lewis2002PRL, deng2019probing, VillegasRosales2021PRB, friess2017negative, bennaceur2018competing, Sun2022}. Recent scanning probe experiments have directly imaged the analogous multi-electron crystalline phases in twisted WS$_2$ moiré superlattices and manifested their rich real-space density profiles~\cite{Li2024Science_WMC}. Previous theoretical studies of Wigner and bubble crystals have mainly focused on zero-temperature energetics based on Hartree-Fock (HF) or density-matrix renormalization group (DMRG) calculations~\cite{maki1983static, Lam, Cote2003, Cooper2003PRL, yoshioka2002dmrg, lewis2004evidence, cote2005replica, Ettouhami2006PRB, cote2008dynamical}, which capture the main features of phase competition and phase diagrams, but do not provide quantitative predictions for $T_m$. Predicting melting in 2D electron bubble phases is not merely a matter of computational complexity, but requires understanding and careful treatment of collective excitations coupled with Landau-level quantization and many-body interactions. Particularly in high LLs, the electron wavefunctions are strongly reshaped by Landau-level form factors, and the resulting spatial separation significantly weakens the short-range Coulomb repulsion and stabilizes charge clustering. Additionally, the presence of impurity potentials in real materials breaks the quasi-long-range order of 2D crystals, meaning the reliability of theoretical calculations must be evaluated through direct comparison with experimental data.

In this Letter, we combine systematic transport measurements of electron bubble phases in high-mobility GaAs/AlGaAs 2DEGs (from LLs $N=2$ to $5$) with theoretical calculations to quantitatively capture the melting process. By applying LL-projected Hartree-Fock elasticity with the full KTHNY melting criterion and its renormalization-group (RG) flow, we for the first time obtained melting temperatures $T_m$ that quantitatively agree with experiment for the bubble phases across different LLs and spin branches. Our results established defect-mediated topological melting as a predictive framework
for correlated quantum Hall electron solids, accurately capturing the underlying connection between microscopic LL physics and macroscopic finite-temperature phase boundaries. In addition, comparison with experiments enables the extraction of topological defect energetics and interaction screening, making the measurement of the solid-liquid transition (SLT) a quantitatively reliable bulk probe of screening and defect physics in quantum Hall solids in the presence of strong electron-electron interaction.

\begin{figure}
\includegraphics[width=0.9\linewidth]{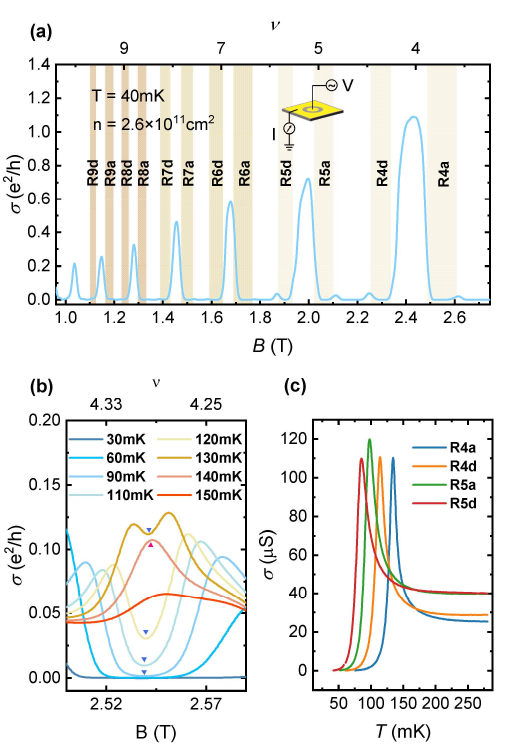}
\centering
\caption{(a) Longitudinal conductance as a function of out-of-plane magnetic field B at T = 40 mK, N=2 to 4 Landau levels ($4 < \nu < 10$). Different bubble states are characterized by previous convention and illustrated with shaded stripes of different colors representing varying Landau levels. (b) Temperature-dependent magnetoresistance around R4a state. Blue triangles mark dips corresponding to the insulating R4a bubble state, the red triangle marks the transition from dip features to peak features, signaling the SLT of R4a state. (c) Longitudinal conductance vs $T$ at filling factors corresponding to maximum transition temperature. Peaks in curves are assigned to be SLT temperatures $T_m$ of bubble phases. }
\label{fig:Fig1}

\end{figure}

\textit{Experiment--}
We perform magnetoresistance measurements on a two-dimensional electron gas (2DEG) located $195\,\mathrm{nm}$ below the surface, confined in a $\delta$-doped $30\,\mathrm{nm}$-wide
$\mathrm{Al}_x\mathrm{Ga}_{1-x}\mathrm{As}/\mathrm{GaAs}/\mathrm{Al}_x\mathrm{Ga}_{1-x}\mathrm{As}$
quantum well ($x=0.24$), with carrier density $n=2.6\times10^{11}\,\mathrm{cm}^{-2}$ and mobility $\mu=2.8\times10^{7}\mathrm{cm}^2/\mathrm{V}\, \mathrm{s}$ , in a dilution
refrigerator. The sample is first characterized in a van der Pauw geometry, where well-developed
reentrant integer quantum Hall (RIQH) states are observed, indicating the emergence of electron
bubble phases. It is then specifically patterned into a Corbino geometry (inset of Fig.~\ref{fig:Fig1}a ;
inner/outer diameters $1.0/1.2\,\mathrm{mm}$) to eliminate the edge-state contributions.

Fig.~\ref{fig:Fig1}a shows the longitudinal conductivity $\sigma$ of the Corbino-geometry sample (in units of the quantum conductance, $e^2/h$) measured at $40\,\text{mK}$ as a function of magnetic field $B$ with corresponding filling factors ($\nu=nh/eB$) indicated on the top axis. The data is dominated by large conductance peaks centered near half-integer filling factors ($i + 1/2$), which originate from emerging stripe-like charge order. The electron bubble phases of interest are identified by the conductivity minima located between these major features and adjacent satellite conductance peaks. For clarity, we adopt the conventional notation~\cite{Deng2012a} $Na$ and $Nd$ (e.g., $4a, 4d$) to label different bubble phases, where the integer $N$ indexes the specific spin-split Landau level branch, and the suffixes $a/d$ denote particle-hole conjugate states within that branch. To probe the evolution and thermal stability of these states, we track the change in magneto-conductivity at different temperatures. We find that the broad conductivity minima initially observed at base temperature gradually narrow and vanish at a certain point. This evolution marks the melting of electron-bubble phases, allowing the extraction of the solid-liquid transition temperature ($T_m$) by identifying peaks in the non-monotonic temperature dependence of the conductivity. We also observe that the thermal stability of bubble phases varies significantly with filling factor; $T_m$ is maximized at an optimal filling factor (indicated by the red triangle in Fig.~\ref{fig:Fig1}b) and decreases away from this point, forming a characteristic dome-like phase diagram.

\begin{figure}
    \centering
    \includegraphics[width=0.85\linewidth]{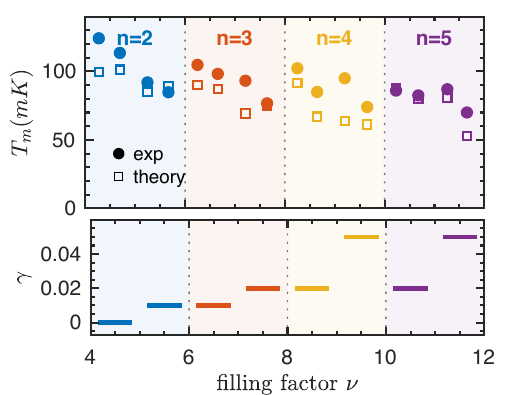}
    \caption{Top panel: $T_m$ versus filling factor $\nu$ in Landau levels $N=2$--$5$. Filled circles show experimental $T_m$ values, and open squares show KTHNY--RG results. Colors label the LL index; shaded area marks the corresponding intervals $\nu\in[2N,2N+2]$. Bottom panel: $\gamma$ used in the calculation, solid (dashed) lines are for the lower (higher) bands in each LL.}

    \label{fig_2}
\end{figure}

As a large melting temperature indicates a more stable phase, we fixed the magnetic field at these filling factors and performed a detailed temperature-dependent conductivity measurement (Fig.~\ref{fig:Fig1}c). The melting temperature ($T_m$) from conductivity peaks is extracted as a function of the partial filling factor ($\nu=\nu_{0}-2N$) at different Landau levels (Fig.~\ref{fig_2}, top panel). With increasing $\nu$ in a given Landau level, $T_m$ roughly follows a smooth decreasing trend, yet exhibits a discrete jump when transitioning to a higher Landau level. This behavior can be qualitatively explained by adapting the universal KTHNY transition picture to the bubble phase. If we treat the bubbles as classical objects with effective charge $Q = Me$ (where $M$ is the number of electrons per bubble), the KTHNY theory yields the melting criterion~\cite{Goerbig2004} $4\pi k_BT_c=0.283e^{2.5}M^{1.5}\sqrt{\frac{B\nu^{*}}{h}}$. Where $\nu^*=\nu \space \text{ mod}\, \frac{1}{2}$ is the partial filling factor defined as the decimal part of filling factor $\nu$. $k_B$ is the Boltzmann constant, $T_c$ is the critical temperature. This equation shows the transition temperature is a function of site electron occupation M, which is theoretically predicted to be dependent on Landau level $N$, with $M=N$ and $M=N+1$ being the most possible configurations. Assuming M=N under the classical picture, we calculate the expected temperature ratios between consecutive Landau levels and compare them with our experimental results. The predicted ratios of 1.49 (for $N=2 \to 3$) and 1.07 (for $N=3 \to 4$) show reasonable agreement with the observed values of 1.35 and 1.17, respectively. 

However, despite this simple classical picture capturing the qualitative inter-level jumps, it remains fundamentally inadequate that the absolute $T_c$ derived from it is severely overestimated. For example, for the N4a phase ($\nu=4.27$), the formula gives $T_c\sim 13K$, far above the measured SLT point $\sim 0.15K$. Moreover, the functional dependence within a Landau level deviates significantly from the prediction. Our data exhibits a nearly linear dependence of $T_m$ on the magnetic field, while classical theory predicts a square-root scaling ($\propto \sqrt{B}$).

\textit{Microscopic model of bubble crystals--}
Motivated by the experimental results above, we develop a theoretical framework for the melting of electron bubble phases. The ground states of a 2DEG in $N \geq 2$ LLs are believed to be insulating electron crystals evolving from a hexagonal (triangular) Wigner Crystal at very small partial filling $\nu^*$ to bubble crystals with an increasing number of electrons $M$ per unit cell as $\nu^*$ increases~\cite{Fogler1996, FoglerKoulakovShklovskii1996PRB, Moessner1996, yoshioka2002dmrg, shibata2001ground, Ettouhami2006PRB, Ro2020}. The total filling factor of a sample under the perpendicular magnetic field $B$ is $\nu = n_e h/eB$, $\nu = \nu^* + 2N$. In the semiclassical model of bubble phases, $M$ electrons cluster at lattice sites ${\mathbf{R}_j}$ to form a bubble, leading to the electron density $n_{N, M}(\mathbf{q}) = \tilde{n}_{N, M}(\mathbf{q})\, \sum_j^{N_b} e^{-i\mathbf{q} \mathbf{R}_j}$,
where $\tilde{n}_{n, M}(\mathbf{q}) = \int d\mathbf{r}\, \sum_{m=0}^{M-1} \ | \phi_{m}^{(n)} (\mathbf{r}) |^2\, e^{-i\mathbf{q}\cdot \mathbf{r}} $ is the density of one bubble at the origin.
$\phi_{m}^{(n)} (\mathbf{r})$ is the single-body wavefunction at the $m_{th}$ Landau orbital in $N$LL (see supplementary materials (SM) for more details \cite{SM}).

The interaction in $N$LL from the Hartree-Fock (HF) calculation is
\begin{equation}
    V_{N,M} = \frac{1}{2} \int \frac{d^2\mathbf{q}}{(2\pi)^2} \left[V_{H}^{(N)}(\mathbf{q}) + V_{F}^{(N)}(\mathbf{q}) \right]\times  |\, \rho_{N,M}(\mathbf{q})\, |^2
    \label{eq_int_HF}
\end{equation}
with $\rho_{N, M}(\mathbf{q}) = n_{N,M}(\mathbf{q})/F_N(\bm q)$, the projected guiding-center density in the $N_{th}$ LL with form factor $F_N(\bm q) = e^{-q^2 \ell_B^2/4}\, \mathcal{L}_N\left( q^2 \ell_B^2/2\right)$ and $q=|\bm q|$, $\ell_B = \sqrt{\hbar/eB·
}$ is the magnetic length. $\mathcal{L}_N\left( x\right)$ is the $N_{th}$ Laguerre polynomial. 
$V_{H}^{(N)}(\mathbf{q})$ and $V_{F}^{(N)}(\mathbf{q})$ are the $N$LL projected Hartree and Fock terms of the interaction, which in the case of Coulomb interaction are:
\begin{align}
    V_{H}^{(N)}(\mathbf{q}) 
    &= \frac{e^2}{\epsilon \ell_B} \left[F_N(q) \right]^2\, \frac{2 \pi}{q\ell_B} \notag \\
    V_{F}^{(N)}(\mathbf{q}) 
    &= -\int d^2 \mathbf{q}'\ V_{H}^{(N)}(\mathbf{q}')\ e^{-i\ell_B^2\, \mathbf{q}' \times \mathbf{q}}
    \label{eq_hartree_fock}
\end{align}
$\epsilon$ is the dielectric constant. Using Poisson's summation in Eq.~\eqref{eq_int_HF}, the cohesive energy (interaction energy \textit{per electron}) of an M-electron bubble crystal is~\cite{SM}:
\begin{equation}
\begin{split}
E_{N,M}
&= \frac{1}{2A_c M} \sum_{\mathbf Q}
\Bigl[ V_H^{(N)}(\mathbf Q)\bigl(1-\delta_{\mathbf Q,0}\bigr)
      + V_F^{(N)}(\mathbf Q) \Bigr] \\
&\qquad \times \bigl| \tilde{\rho}_{N,M}(\mathbf Q) \bigr|^2
\end{split}\label{eq_E}
\end{equation}
$\mathbf{Q}$ runs over all the reciprocal lattice vectors and $A_c = 2\pi \ell_B^2\,  M/\nu^*$ is the unit area of this crystal. The projected one-bubble guiding-center density $\tilde{\rho}_{N,M}(\mathbf{q})$ is defined as $\tilde{n}_{N,M}/F_N(\bm q)$. In high LLs, the screening effect from lower filled LLs generally cannot be ignored and can be encoded in the dielectric constant $\epsilon (q)$ by Aleiner-Glazman's empirical equation: $\epsilon (q) = \epsilon \left[ 1 + {2\gamma} \left[ 1-J_0^2 (q R_c) \right]/(q a_B) \right]$~\cite{AleinerGlazman1995PRB, FoglerKoulakovShklovskii1996PRB}. $a_B = \hbar^2 \epsilon / m^*e^2$ is the effective Bohr radius ($m^*$ is the effective mass), and $R_c = \sqrt{2N+1}\, \ell_B$ is the cyclotron radius of one electron in $N$LL. Screening strength is given by $\gamma$.

From the energy of a 2D crystal, one can extract its shear modulus $\mu$ and the second Láme coefficient $\lambda$ using classical elastic theory. We first introduce a stiffness tensor, dynamical matrix $D_{\alpha \beta}(\mathbf{q})$ with $\alpha,\, \beta = x,\, y$, that represents the restoring force produced from a distortion in lattices. For a 2D Bravais lattice with pairwise interaction $V(\mathbf{q})$ (such as Coulomb) and unit area $A_c$, 
\begin{equation}
\begin{split}
    D_{\alpha \beta}(\mathbf{q}) = \frac{1}{A_c} \sum_{\mathbf{Q}} 
    &(q_\alpha+Q_\alpha)(q_\beta+Q_\beta)\, V(\mathbf{q} + \mathbf{Q}) \\
    & - Q_\alpha Q_\beta\, V(\mathbf{Q})
\end{split}
   \label{eq_dynamic_matrix_momentum}
\end{equation}
In the long wavelength limit, the dynamical matrix of an isotropic triangular lattice under the harmonic approximation is related to the elastic constants by
\begin{equation}
    D_{\alpha\beta} (\mathbf{q}) = ( \mu + \lambda)\, q_\alpha q_\beta + \mu\, q^2 \delta_{\alpha, \beta}
    \label{eq_D}
\end{equation}
Expanding Eq. \eqref{eq_dynamic_matrix_momentum} up to the second order of $V$ and using Eq. \eqref{eq_D}, the two elastic constants can be expressed by the interaction as:
\begin{equation}
    \mu = \frac{\nu^*}{4\pi l_B^2 M} \sum_{\mathbf{Q}\ne 0} Q_x^2\Bigg\{ \frac{V'(Q)}{Q} + Q_y^2  \frac{QV''(Q)-V'(Q)}{Q^3} \Bigg\}
    \label{eq_mu}
\end{equation}
\begin{equation}
\begin{split}
    \lambda = \frac{\nu^*}{2\pi l_B^2 M} &\sum_{\mathbf{Q}\ne 0} \Bigg\{ V(Q) + \frac{3}{2} Q_x^2 \frac{V'(Q)}{Q} \\
    &+  Q_x^2 \left(\frac{Q_x^2}{2}-Q_y^2 \right) \frac{QV''(Q)-V'(Q)}{Q^3}\Bigg\}
    \label{eq_lambda}
\end{split}     
\end{equation}

To make connections to the melting temperature $T_m$, we introduce two more quantities. The 2D Young's modulus is defined as $Y \equiv {4\mu (\mu+\lambda)}/({2\mu+\lambda})$ from which a dimensionless coupling constant $K$, the \emph{2D Cauchy modulus}, can be introduced: $K = {A_c\, Y}/({k_B T})$, where $T$ denotes temperature. The universal criterion of a 2D solid-liquid transition (SLT) is $K = 16\pi$. Therefore, the first natural estimation of $T_m$ is using $\mu_0$ and $\lambda_0$ from Eq.~\eqref{eq_mu} and~\eqref{eq_lambda} to derive:
\begin{equation}
    T_m = \frac{A_c \cdot Y_0}{16\pi k_B}, \qquad Y_0 = \frac{4\mu_0 (\mu_0 + \lambda_0)}{2\mu_0 +\lambda_0}
    \label{eq_t_0}
\end{equation}
This estimate, however, has serious limitations for using the bare elastic constants obtained from the zero-temperature HF directly, since the melting is a finite temperature process and thermal fluctuations strongly soften the lattice. As a result, Eq.~\eqref{eq_t_0} systematically overestimates the true $T_m$ and should be regarded only as an upper bound or a rough estimate.

\textit{KTHNY melting with renormalization group refinement--}
It is important to note that the melting of a 2D quasi-long-range order is driven by thermally excited \emph{topological defects}~\cite{Kosterlitz1973, kosterlitz1973long, kosterlitz1978two}. A crystal at low temperatures only hosts tightly bound dislocation pairs with opposite Burgers vectors $\mathbf{b}$. As the temperature rises, thermal fluctuations will gradually weaken the attraction inside the pairs until they unbind and proliferate, during which the elastic moduli will be softened~\cite{halperin1978theory, nelson1979dislocation, mermin1979topological, young1979melting}. We adopt the KTHNY framework with the renormalization-group (RG) treatment to consistently account for this defect-mediated softening.

 The dislocation core energy is approximated as $E_d \approx \alpha A_c Y$, $\alpha$ is a constant that controls the energetic cost of creating a dislocation. Dislocation fugacity $y(T)\equiv \text{exp} (-E_c/k_BT) = \text{exp} \left(-\alpha K \right)$ measures the Boltzmann weight for thermally exciting a dislocation core. We use $l = \ln |\textbf{b}|$ as the logarithm of the coarse-graining length in units of the microscopic cutoff. $l=0$ corresponds to the unrenormalized values, while $l\rightarrow \infty$ yielding the renormalized ones. The coupled flow equations for a triangular lattice are:
\begin{equation}
    \frac{d}{dl} \left( \frac{1}{K(l)} \right) = A(K)\, y(l)^2
    \label{eq_rg_k}
\end{equation}
\begin{equation}    
    \frac{d\, y(l)}{d\, l} = \left( 2- \frac{K(l)}{8\pi} \right)\,y(l) + B(K)\, y(l)^2
    \label{eq_rg_y}
\end{equation}
$A(K) =  \frac{3\pi}{4} e^{K(l)/8\pi} \{ 2 I_0\left(K(l)/8\pi \right) - I_1\left(K(l)/8\pi \right) \}$, and $B(K) = 2\pi e^{K(l)/16\pi}\ I_0\left(K(l)/8\pi \right)$; $I_{a}(x)$ are modified Bessel functions. Eq.~\eqref{eq_rg_k} and \eqref{eq_rg_y} feature a separatrix terminating at the critical melting point, where the renormalized elastic coupling satisfies: $K_R(T_m^-)=16\pi, T_m$ is obtained by integrating the coupled RG flow equations with $l\rightarrow\infty$. At any temperature, the renormalized 2D Cauchy modulus is $K_R(T) \equiv \lim_{l\to\infty} K(l)$. The initial conditions are fixed by the elastic moduli $(\mu_0,\lambda_0)$ from the zero-temperature HF computations.

\textit{Discussions--} 
The melting temperature $T_m$ computed from the KTHNY-RG framework is compared with the experimental melting temperatures with the two tuning parameters $\alpha$ and $\gamma$. The resulting $T_m$ values are summarized in Figure~\ref{fig_2}, together with the extracted screening strength $\gamma$ across LLs and spin branches. A larger $\gamma$ implies a weaker effective interaction within the partially filled Landau level, therefore reduces the bare elastic moduli, softens the bubble lattice, and consequently lowers $T_m$. Within each Landau level, the fitted $\gamma$ is consistently larger for the upper spin branch than for the lower one. This trend is consistent with the much smaller spin splitting compared with the cyclotron gap, which strengthens screening for the upper branch.

\begin{figure}
    \centering
    \includegraphics[width=0.9\linewidth]{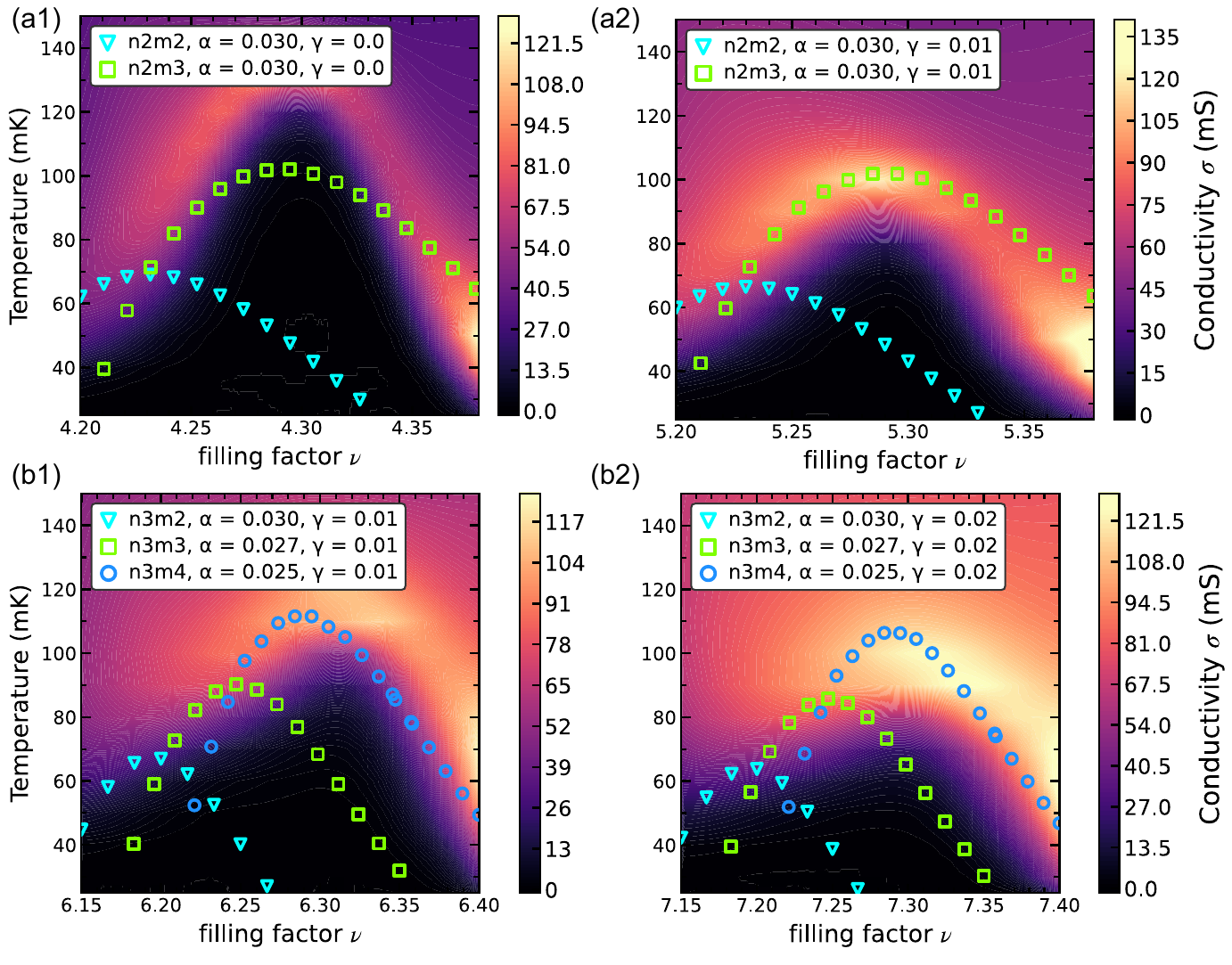}
    \caption{Heat maps of the longitudinal conductance as a function of temperature $T$ and filling factor $\nu$. Overlaid symbols mark the melting temperatures $T_m(\nu)$ for the indicated $M$-electron bubble phases obtained from theory ($\alpha$ and $\gamma$ are listed in the legends). (a) for $2$LL, (b) for $3$LL.}
    \label{fig_3}
\end{figure}

Figure~\ref{fig_3} presents temperature-dependent conductivity in the $N=2\sim 3$ LLs as heat maps for both spin branches (see~\cite{SM} for other LLs). The overlaid scatter points show $T_m (\nu^*)$ from the KTHNY–RG calculation for the candidate $(N, M)$ bubble crystals, using the fitted $\alpha$ and $\gamma$ indicated in each panel. The quantitative agreement between theory and experiment supports the identification of the reentrant insulating regions as bubble crystals and is consistent with a gradual increase of the electron number $M$ per bubble with filling factor. Crossings between $M$- and $(M+1)$-bubble curves mark the fillings where the preferred bubble occupancy changes theoretically, indicating structural transitions between competing bubble lattices. The non-monotonic dependence of $T_m$ and its systematic evolution with $M$ across panels verifies a defect-mediated melting mechanism. 

The fitted values of $\alpha$ are consistently small across different bubble phases, indicating a low dislocation core energy, meaning that topological defects are easily created at relatively low temperatures. Consequently, dislocations strongly soften the lattice and renormalize $\mu$ and $\lambda$, explaining why a bare zero-temperature estimate fails to reproduce the measured $T_m$. In this sense, a smaller $\alpha$ also signifies a more fragile electronic solid, whose rigidity is more readily undermined by defect proliferation and thermal fluctuations.

In conclusion, we perform bulk Corbino transport measurements that track the evolution of electron bubble phases across a wide range of Landau levels, and develop a theoretical framework that quantitatively captures their melting temperatures. Beyond identifying the conductivity and thus $T_m$ of a 2D electronic crystal, the comparison between experiment and theory allows us to extract the dislocation-core parameter $\alpha$ and the screening strength $\gamma$ in the relevant Landau level. Here $\alpha$ quantifies the relative importance of dislocation core energetics versus long-wavelength elasticity (i.e., the fragility of the electronic solid to fluctuation-driven phase transition), while $\gamma$ calibrates the effective interaction, thereby constraining the competing electronic phases as well as the energy scales of their collective modes and excitations. The experimentally obtained SLT boundary, therefore, provides a bulk constraint on the microscopic cost of creating topological defects and the dressing of interaction. 

Our strategy can be extended to include disorder, anisotropy, finite thickness, and Landau-level mixing. More broadly, it provides a route to study the melting and phase competition in other 2D electronic crystals, including solids proximate to fractional quantum Hall liquids, where the fundamental effective particles are emergent quasiparticles (anyons). Understanding the existence of solid phases there is essential for studying the boundaries of topological order. It also applies to the recently observed generalized Wigner and multi-electron crystals in TMD moiré systems~\cite{Li2021Nature_GWCimaging, padhi2021generalized, huang2021correlated, Li2024Science_WMC}, where the required energetics and elastic response can be computed from band-projected form factors and interactions, enabling a unified description of flat-band screening, interaction dressing, and the stability of charge-ordered phases in Chern-band systems.

\textit{Acknowledgement--} We thank Yuan Li, Junren Shi, and Ruirui Du for discussions. The authors from Peking University acknowledge support by the National Key Research and Development Program of China (Grant No. 2021YFA1401900). The authors from Nanyang Technological University are supported by the National Research Foundation, Singapore, under the NRF Fellowship Award (NRF-NRFF12-2020-005), Singapore Ministry of Education (MOE) Academic Research Fund Tier 3 Grant (No. MOE-MOET32023-0003) “Quantum Geometric Advantage”, and Singapore Ministry of Education (MOE) Academic Research Fund Tier 2 Grant (No. MOE-T2EP50124-0017). This work is supported by the Quantum Science and Technology--National Science and Technology Major Project (Grant No. 2021ZD0302600). The Princeton University portion of this research is funded in part by the Gordon and Betty Moore Foundation's EPiQS Initiative, Grant GBMF9615.01 to Loren Pfeiffer.

\bibliography{cleaned_bib_v1.1}

\clearpage
\onecolumngrid

\setcounter{equation}{0}
\setcounter{figure}{0}
\setcounter{table}{0}
\setcounter{section}{0}
\renewcommand{\theequation}{S\arabic{equation}}
\renewcommand{\thefigure}{S\arabic{figure}}
\renewcommand{\thetable}{S\arabic{table}}

\begin{center}
  \textbf{\large Supplemental Material for\\ “Melting of quantum Hall Wigner and bubble crystals"}
\end{center}
\vspace{1em}

This Supplemental Material provides additional details and derivations supporting the main text. We summarize the physical picture and notation for quantum Hall Wigner/bubble crystals, present the Landau-level--projected Hartree--Fock formulation of bubble-crystal density profiles and cohesive energies (including the modeling with screening/finite-thickness), and derive the elastic moduli for 2D triangular lattices from the microscopic interaction kernel. We further outline the defect-mediated KTHNY melting framework and the associated renormalization-group flow equations used to obtain the melting temperature \(T_m\).


\section{Physical Picture}
Our system is 2DEG in a perpendicular magnetic field $B\hat{\mathbf{z}}$. The magnetic flux $\mathbf{\Phi}$ is quantized and defined as:
\begin{equation}
    \Phi = \mathbf{B}\cdot \mathbf{S} = N_\phi \Phi_0,\ N_\phi = 0, 1, 2,...
\end{equation}
where $\Phi_0 = \frac{h}{e}$ is the magnetic flux quanta and $\mathbf{S}$ being the surface encircling the magnetic fluxes. In our case, $\mathbf{B}\perp \mathbf{S}$. $N_\phi = BS/\Phi_0$ is the number of flux quanta. Filling factor $\nu$ is defined as:
\begin{equation}
    \nu \equiv \frac{N_e}{N_\phi} = \frac{N_e}{BS}\Phi_0 = n_e\frac{h}{eB} = n_e \frac{2\pi \hbar}{eB} = n_e\cdot 2\pi \ell_B^2
\end{equation}
$n_e= N_e/S$ is the average electron density in the system, fixed in the experiment.
Magnetic length $l_B$ serves as the fundamental length scale in the system
\begin{equation}
    \ell_B \equiv \sqrt{\frac{\hbar}{eB}} = \sqrt{\frac{\nu}{2\pi n_e}}
\end{equation}

In $N>1$ LLs, the ground state of the electron system is the pinning mode (insulating phase) that evolves from a hexagonal Wigner Crystal to bubble crystals with an increasing number of electrons $M$ per lattice site as partial filling $\nu^*$ increases. Mean field theory predicts the final bubble state to be $M=N+1$, while in DMRG, the last bubble state is unstable, and a transition from $M=N$ to stripe phase happens directly when $\nu^* \approx 1/2$. Electrons in the filled $N-1$ LLs below the $N$ LL are taken to be completely inert.
\begin{equation}
    \nu = \nu^* + 2N, \ 0<\nu^*<2
\end{equation} 
We usually constrain $0<\nu^*<1/2$, other parts can be accounted for by particle-hole symmetry and different spin branches.

Number of the electrons on the topmost unfilled LL $N_e^* = n_e^* \cdot S$. These electrons form bubbles and satisfy $N_e^* = M\cdot N_b$
where $N_b$ is the number of bubbles. Using $A_c$ to denote the unit area for the bubble lattice, $A_c N_b =S$. The unit area of the bubble lattice is
\begin{equation}
    A_c = 2\pi l_B^2 \left( \frac{M}{\nu^*} \right)
\end{equation}
For a hexagonal lattice, $A_c = \sqrt{3}a^2/{2} $
gives the lattice constant of the bubble crystal $a$:
\begin{equation}
    a =  \sqrt{\frac{4\pi M}{\sqrt{3} \nu^*}} l_B = \sqrt{ \frac{2M}{\sqrt{3} n_e} \left( \frac{2N}{\nu^*} +1 \right) }
\end{equation}

A bubble appears when the guiding centers of $M$ electrons accumulate. So it can be regarded as a guiding center density wave. Cyclotron radius of a single electron at $N$ LL (computed from the density):
\begin{equation}
    R_c = \sqrt{2N+1}\ l_B
\end{equation}
serves as another fundamental length scale (also the largest one) at high LLs besides the magnetic length itself. The aggregation of $M$ particles in size domains at the order of $R_c$ allows the system to achieve a lower exchange energy, namely the Fock term more dominant than the Hartree term in higher LLs.

\section{Microscopic model of bubble crystals}
For a bubble crystal at $N$LL with $M$ electrons per bubble, the real-space density is usually approximated to be
\begin{equation}
    n_{N,M}(\mathbf{r}) = \sum_j^{N_b} \sum_{m=0}^{M-1} | \phi_{m}^{(N)} (\mathbf{r}-\mathbf{R}_j) |^2
\end{equation}
under the assumption that all the bubbles are identical, where $\phi_{m}^{(N)} (\mathbf{r})$ is the single particle wavefunction in $N$LL with Landau orbital $m = 0, 1, 2, ...$, which is given by
\begin{equation}
    \phi_{m}^{(N)} (\mathbf{r}) = C_{m}^{(N)} \left(\frac{r}{\ell_B} \right)^{|m-N|} e^{-r^2/4\ell_B^2} e^{i(N-m)\theta}\ L_{(N+m-|m-n|)/2}^{|m-N|} (r^2/2\ell_B^2)
    \label{eq_wf_single}
\end{equation}
under symmetric gauge $\mathbf{A} = (-yB/2, xB/2)$.

The density of one bubble located at the origin $\tilde{n}_{N, M}(\mathbf{r})$ is:
\begin{equation}
    \tilde{n}_{N,M}(\mathbf{r}) = \sum_{m=0}^{M-1} | \phi_{m}^{(N)} (\mathbf{r}) |^2
    \label{eq_density_real_single}
\end{equation}
Its Fourier transform in momentum space is:
\begin{align}
    \tilde{n}_{N,M}(\mathbf{q}) &= \int d\mathbf{r} \sum_{m=0}^{M-1} \ | \phi_{m}^{(N)} (\mathbf{r}) |^2 e^{-i\mathbf{q}\cdot \mathbf{r}}
\end{align}
Using Eq. \eqref{eq_wf_single}, it can be further simplified for an isotropic system as
\begin{equation*}
    \tilde{n}_{N,M}(\mathbf{q}) = \sum_{m=0}^{M-1}\ \tilde{C}_m^{(N)} \int_0^\infty dx \cdot e^{-x} \cdot x^{|m-N|}\cdot \left[ \mathcal{L}_{(n+m-|m-N|)/2}^{|m-N|} (x) \right]^2 J_0(\sqrt{2x}\cdot q \ell_B)
\end{equation*}
\begin{equation}
    \tilde{C}_{m}^{(N)} = \frac{\text{min}(N!, m!)}{\text{max}(N!, m!)}
\end{equation}
This is the density of one bubble in momentum space.

\section{Cohesive energies from HF computation with (screened) Coulomb interactions}
In this section, we introduce the zero-temperature Hartree-Fock description of the bubble crystal, as well as the model after considering the screening and finite-thickness effects.

\subsection{Hartree-Fock cohesive energy}
The Hartree-Fock interaction of realistic Coulomb interaction in $n_{th}$ LL is given by:
\begin{equation}
    V(\mathbf{q})^{(N)} = \frac{1}{2} \int \frac{d^2\mathbf{q}}{(2\pi)^2} [V_{H}^{(N)}(\mathbf{q}) + V_{F}^{(N)}(\mathbf{q})]\cdot  |\langle \hat{\rho}_{N,M}(\mathbf{q})\rangle|^2
    \label{eq_int_HF}
\end{equation}
where the Hartree and Fock terms are:
\begin{align}
    V_{H}^{(N)}(\mathbf{q}) &= \frac{e^2}{\epsilon l_B} \frac{2 \pi}{q} [F_N(q)]^2 =  \frac{e^2}{\epsilon l_B} \frac{2 \pi}{q} e^{-q^2/2} \left[ \mathcal{L}_N\left( \frac{q^2}{2}\right) \right]^2 \label{eq_hartree} \\
    V_{F}^{(N)}(\mathbf{q}) &= -\int d \mathbf{q'}\ V_{H}^{(N)}(\mathbf{q'})\ e^{-i \mathbf{q'} \times \mathbf{q}} \\
    &= -  \frac{e^2}{\epsilon l_B} 2\pi
    \int_0^\infty dq' \cdot [F_n(q')]^2 J_0(q \cdot q')\notag \\
    &= -  \frac{e^2}{\epsilon l_B} 2\pi
    \int_0^\infty dq' \cdot e^{-q'^2 /2} \left[\mathcal{L}_n \left( \frac{q'^2}{2} \right) \right]^2 J_0(q\cdot q') \label{eq_fock}
\end{align}
$\hat{\rho}(\mathbf{q})$ is the projected electronic density $\hat{n}(\mathbf{q})$ onto the $n_{th}$ LL. I will use $\rho(\mathbf{q})$ to represent  $\langle {\Psi}| \hat{\rho}(\mathbf{q}) |{\Psi} \rangle$ for simplicity.
\begin{equation}
    \rho(\mathbf{q}) = \frac{n_{n,M}(\mathbf{q})}{F_n(q)}
\end{equation}

We can write the projected density as:
\begin{equation}
    \rho_{n,M}(\mathbf{q}) = \frac{\tilde{n}_{n,M}(\mathbf{q})}{F_n(q)}\ \sum_j^{N_b} e^{-i\mathbf{q} \mathbf{R}_j} \equiv \tilde{\rho}_{n,M}(\mathbf{q}) \sum_j^{N_b} e^{-i\mathbf{q} \mathbf{R}_j}
\end{equation}
where $\tilde{\rho}_{n,M}(\mathbf{q})$ means the projected one-bubble density to $n_{th}$ LL at the origin in momentum space. Substituting the equation into Eq. \eqref{eq_int_HF}, we get the expression for cohesive energy (the Hartree-Fock energy \textit{per electron} in the partially filled level):
\begin{equation}
    E_{n,M} = \frac{1}{2M\cdot A} \sum_{\mathbf{Q}} \left[ V_H(\mathbf{Q}) (1-\delta_{\mathbf{Q},0}) + V_F(\mathbf{Q}) \right]\cdot | \tilde{\rho}_{n,M}(\mathbf{Q})|^2
\end{equation}
The projected density is explicitly given by:
\begin{equation}
    \tilde{\rho}_{n,M}(\mathbf{q})
    = \frac{1}{F_n(q)} \sum_{m=0}^{M-1}\ \tilde{C}_m^{(n)} \int_0^\infty dq' \cdot e^{-q'} J_0(\sqrt{2q'}\cdot ql_B)\cdot q'^{|m-n|}\cdot [\mathcal{L}_{(n+m-|m-n|)/2}^{|m-n|} (q')]^2
\end{equation}

\subsection{Effects of screening of finite thickness}

The effect of lower LLs is encoded in the dielectric constant $\epsilon(q)$ from Aleiner-Glazman's empirical equation:
\begin{equation}
    \epsilon (q) = \epsilon \left( 1 + \frac{2\gamma}{q a_B} \left[ 1-J_0^2 (q R_c) \right] \right)
\end{equation}
\begin{equation}
    a_B = \hbar^2 \epsilon / m^*e^2
\end{equation}
$a_B$ is the effective Bohr radius and $R_c = \sqrt{2n+1} l_B$. Typically, the effective mass in the host semiconductor -- GaAs -- is given by $m^* = 0.067 m_e$.

\begin{equation}
    V(q) = \frac{2\pi e^2}{q\epsilon l_B} \frac{e^{-\beta q}}{ 1 + \frac{2 \gamma}{q a_B} \left[ 1-J_0^2 (q R_c) \right] }
\end{equation}
 Define the factor:
\begin{equation}
    S(q,n, \alpha, \gamma) = \frac{e^{-\beta q}}{ 1 + \frac{2 \gamma}{q a_B} \left[ 1-J_0^2 (q R_c) \right] }
\end{equation}
The Hartree-Fock terms simply change to:
\begin{align}
    V_{H}^{(n)}(\mathbf{q}) &= \frac{e^2}{\epsilon l_B } \frac{2\pi}{q} [F_n(q)]^2 \frac{e^{-\beta q}}{ 1 + \frac{2\gamma}{q a_B} \left[ 1-J_0^2 (q R_c) \right] } \label{eq_hartree_screening} = \frac{e^2}{\epsilon l_B } \frac{2\pi}{q} [F_n(q)]^2 \cdot S(q,n, \alpha, \lambda) \\
    V_{F}^{(n)}(\mathbf{q}) &= -\int d \mathbf{q'}\ V_{H}^{(n)}(\mathbf{x})\ e^{-i \mathbf{q'} \times \mathbf{q}} = -  \frac{e^2}{\epsilon l_B} 2\pi 
    \int_0^\infty dq'\ e^{-q'^2 /2} S(q',n, \alpha, \gamma) \left[\mathcal{L}_n \left( \frac{q'^2}{2} \right) \right]^2 J_0(qq') \\
    &= -  \frac{e^2}{\epsilon l_B} 2\pi 
    \int_0^\infty dq' \frac{e^{-\beta q'} \cdot e^{-q'^2 /2} }{ 1 + \frac{2\gamma}{q' a_B} \left[ 1-J_0^2 (q' R_c) \right] } \left[\mathcal{L}_n \left( \frac{q'^2}{2} \right) \right]^2 J_0(qq') \label{eq_fock_screening}
\end{align}
$\gamma$ is generally taken to be the order of $0 \sim 1$ (in unit of $l_B$). $d = \beta l_B$ is the thickness of the quantum well in the Fang-Howard potential. In this work, we only consider the effect of screening $\gamma$, and do not take the finite thickness $\beta$ into account ($\beta = 0$ for all the simulations).

\begin{figure}
    \centering
    \includegraphics[width=0.8\linewidth]{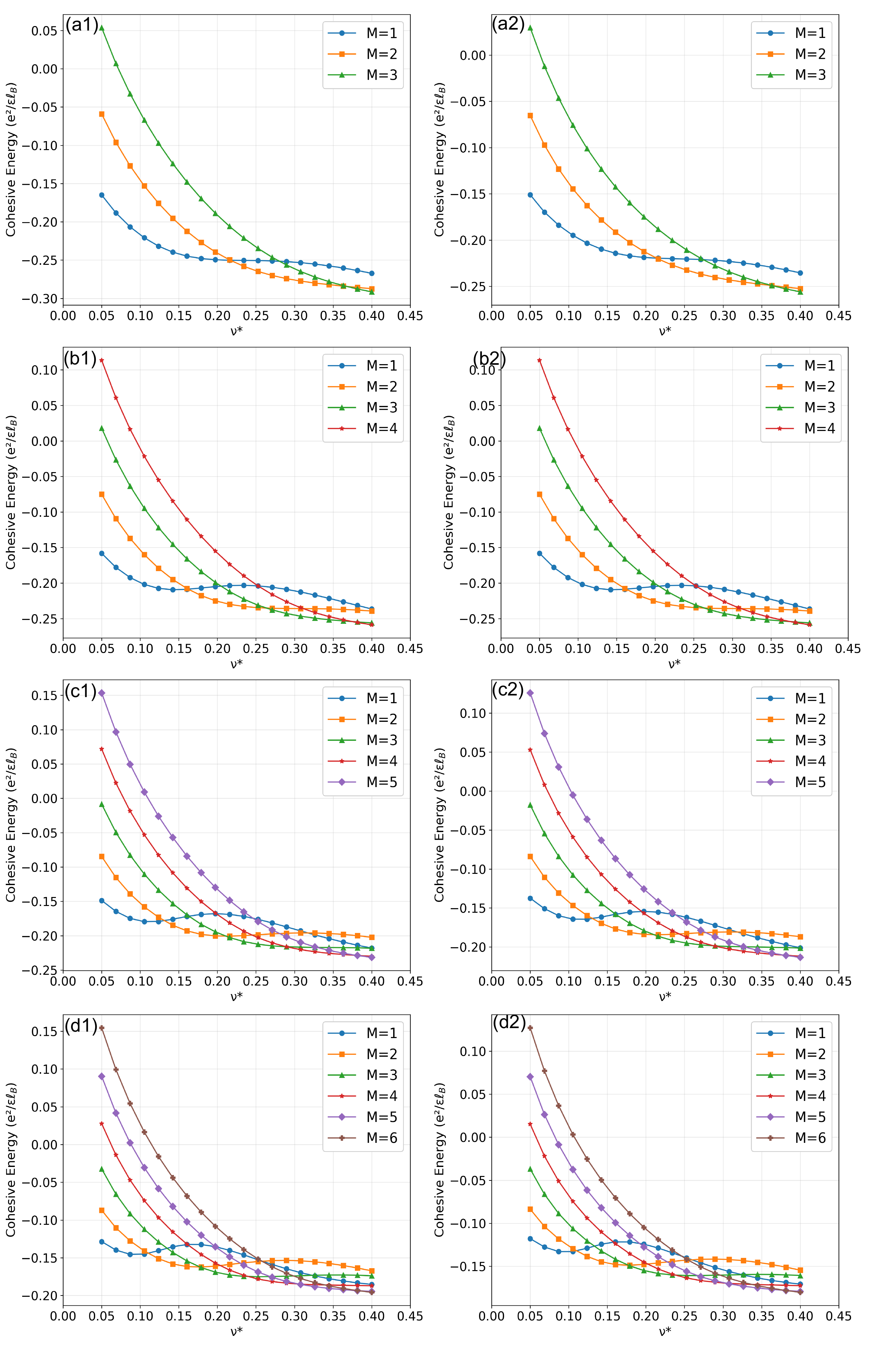}
    \caption{Hartree--Fock cohesive energy of bubble crystals, \(E_{\rm coh}(\nu^\ast)\), as a function of partial filling \(\nu^\ast\) for Landau levels \(n=2\) ((a1),(a2)), \(n=3\) ((b1),(b2)), \(n=4\) ((c1),(c2)), and \(n=5\) ((d1),(d2)). Each curve corresponds to an \(M\)-electron bubble phase (triangular lattice), as labeled in the legends. Calculations are performed with \(\beta=0\) and the screening parameter \(\gamma\) below: (a1)$=0.00$, (a2)$=0.01$, (b1)$=0.01$, (b2)$=0.02$, (c1)$=0.02$, (c2)$=0.05$, (d1)$=0.02$, (d2)$=0.05$. For each \(\nu^\ast\), the lowest \(E_{\rm coh}\) identifies the energetically favored electron-solid morphology at zero temperature.}

    \label{fig_s_E}
\end{figure}

Fig. \ref{fig_s_E} shows the cohesive energies from Hartree-Fock computation at zero temperatures for LLs $n=2\sim5$ with screening parameters the same values as we used in fitting the melting temperature in the main content.

\section{Elastic moduli for 2D triangular lattices}
In this section, we first give the general elastic theory and notations for 2D lattices then give the results for the triangular lattice specifically.

\subsection{Potential energy of a 2D lattice}
Lattice points formed by electrons is $\mathbf{R}$, the displacement field away from the equilibrium $\mathbf{R}$ is $\mathbf{u}(\mathbf{R})$, use $i, j$ to denote the lattice points (electron indices). The total potential energy (from electron-electron interaction) of the crystal in unit of $e^2/\epsilon l_B$ is given by:
\begin{align}
    &U = \frac{1}{2} \sum_{i,j} U(\mathbf{r}_i - \mathbf{r}_j) 
    = \frac{1}{2} \sum_{i,j} U_{ij}, \ \ \mathbf{r}_{i} = \mathbf{R}_i + \mathbf{u}(\mathbf{R}_i)
    \label{eq_potential_general}
\end{align}
$U_{ij}$ is introduced for notification simplicity.

The displacement is very small compared to the lattice separation, $U_{ij}$ can be expanded around $\mathbf{R}_{ij}$. Using the general Taylor expansion of a function with argument $\mathbf{r}$ around its equilibrium value at $\mathbf{r}_0$, the pair energy potential $U_{ij}$ turns out to be:
\begin{align}
    U_{ij} &= U(\mathbf{R}_{ij}) 
    + \delta \mathbf{u}_{ij} \cdot \nabla U(\mathbf{R}_{ij}) 
    + \frac{1}{2!} [\delta \mathbf{u}_{ij} \cdot \nabla]^2 U(\mathbf{R}_{ij}) 
    + ... + \frac{1}{n!} [ \delta \mathbf{u}_{ij} \cdot \nabla] ^n U(\mathbf{R}_{ij}) + ... \\
    U_{ij} &= U_{ij}^{(0)} + U_{ij}^{(1)} + U_{ij}^{(2)} + ... + U_{ij}^{(n)}\\
    \mathbf{R}_{ij} &\equiv \mathbf{R}_{i} - \mathbf{R}_{j}, \ \ \delta \mathbf{u}_{ij} \equiv \mathbf{u}(\mathbf{R}_{i}) - \mathbf{u}(\mathbf{R}_{j})
\end{align}
The coefficient of $\mathbf{u}_{ij}$'s linear term $\sum_{i,j} \nabla U(\mathbf{R}_{i}-\mathbf{R}_{j})$ is $0$, which is the summation of all the negative forces from other particles acting on the object located at $\mathbf{R}_i$. So in the simplest model, the expansion is saved up to $U_{ij}^{(2)}$ and called the simple harmonic approximation (SHA).

Under SHA,
\begin{align}
U &\approx \frac{1}{2} \sum_{i,j} U(\mathbf{R}_{ij}) +  U^{(2)}  \\
U^{(2)} 
&= \frac{1}{4} \sum_{i, j} \sum_{\alpha, \beta}\ \delta u_{ij}^\alpha \frac{\partial}{\partial R^{\alpha}_{ij}} \delta u_{ij}^\beta \frac{\partial}{\partial R^{\beta}_{ij}}\ U(\mathbf{R}_{ij}),\ \alpha, \beta = x,y 
\end{align}

\subsection{Dynamical matrix}
The quadratic term in elasticity theory is usually written in another form by the so-called dynamical matrix, as will be explained below. Define:
\begin{align}
U_{\alpha \beta}\ (\mathbf{R}_{ij}) &\equiv \frac{\partial^2 U(\mathbf{R}_{ij})}{\partial R^{\alpha}_{ij}\ \partial R^{\beta}_{ij}}\\
U^{(2)} &= 
\frac{1}{4} \sum_{i, j} \sum_{\alpha, \beta} 
\left[ u^\alpha(\mathbf{R}_i) - u^\alpha(\mathbf{R}_j) \right]\ U^{\alpha \beta} (\mathbf{R}_{ij}) \left[ u^\beta(\mathbf{R}_i) - u^\beta(\mathbf{R}_j) \right]\\
&= \frac{1}{2} \sum_{i, j} \sum_{\alpha, \beta}
u^\alpha(\mathbf{R}_i) 
D_{\alpha \beta} (\mathbf{R}_{ij}) u^\beta(\mathbf{R}_j) \\
D_{\alpha \beta} (\mathbf{R}_{ij}) 
&= D_{\alpha \beta} (\mathbf{R}_{i}-\mathbf{R}_{j}) 
\equiv 
\sum_{k} U_{\alpha \beta}(\mathbf{R}_{ik}) \delta_{i,j} - U_{\alpha \beta}(\mathbf{R}_{ij}),\ \alpha, \beta = x,y
\label{eq_D(R)}\\
U^{(2)} &= \frac{1}{2} \sum_{i, j} 
\mathbf{u}(\mathbf{R}_i)\ \mathbf{D} (\mathbf{R}_{ij})\ \mathbf{u}(\mathbf{R}_j)
\end{align}
$D_{\alpha \beta} (\mathbf{R}_{ij})$ is the ${{\alpha \beta}}$ term of the $2\times2$ matrix $\mathbf{D}$, often called elastic matrix. The first term of Eq \eqref{eq_D(R)} is also called the on-site terms, while the second is the off-site terms. 

Making a transformation $\mathbf{R} = \mathbf{R}_i - \mathbf{R}_j$, $D_{\alpha \beta} (\mathbf{R}_{ij})$ can be written as:
\begin{equation}
    D_{\alpha \beta}(\mathbf{R}) = \delta_{\mathbf{R}, 0} \sum_{\mathbf{R}'} U^{\alpha \beta}(\mathbf{R}') - 
    U_{\alpha \beta}(\mathbf{R})
    \label{eq_elastic_matrix_R}
\end{equation}
Fourier transform according to the definition:
\begin{align}
D_{\alpha \beta}(\mathbf{q}) &= \sum_{\mathbf{R}} D_{\alpha \beta}(\mathbf{R}) e^{-i \mathbf{q} \cdot \mathbf{R}}\\[3pt]
D_{\alpha \beta}(\mathbf{R}) &= \frac{1}{A_c} \int_{\mathbf{q}} D_{\alpha \beta}(\mathbf{q}) e^{i \mathbf{q} \cdot \mathbf{R}}
\end{align}
where $A_c$ is the area of the primitive cell,
\begin{equation}
    D_{\alpha \beta}(\mathbf{q}) 
    = \sum_{\mathbf{R}} 
    U_{\alpha \beta}(\mathbf{R})
    \left(
    1- e^{-i \mathbf{q} \cdot \mathbf{R}}
    \right)
\end{equation}

Now consider the Fourier transform of the second derivative of the potential energy:
\begin{equation}
    U_{\alpha \beta}(\mathbf{q}) = -q_\alpha q_\beta \int d\mathbf{r} U(\mathbf{r}) e^{-i \mathbf{q} \cdot \mathbf{r}} = -q_\alpha q_\beta U(\mathbf{q})
\end{equation}
where integration by parts and the assumption that the boundary term vanishes are used. The Fourier transform, in which $A_c$ is the area of the unit cell, is given by:
\begin{align}
    U_{\alpha \beta}(\mathbf{R}) &= \frac{1}{A} \int_{\mathbf{q}} U_{\alpha \beta}(\mathbf{q}) e^{i \mathbf{q} \cdot \mathbf{R}}
\end{align}
In a lattice system, use Poisson summation:
\begin{equation}
    \sum_{\mathbf{R}_i} e^{-i \mathbf{q} \cdot \mathbf{R}_i} = \frac{2\pi}{A} \sum_{\mathbf{Q}} \delta(\mathbf{q}-\mathbf{Q})
\end{equation}
\begin{align}
    D_{\alpha \beta}(\mathbf{q})
    &= \frac{1}{A} \sum_{\mathbf{Q}} 
    \left[
    (q_\alpha+Q_\alpha)(q_\beta+Q_\beta) U(\mathbf{q} + \mathbf{Q})
    - Q_\alpha Q_\beta U(\mathbf{Q})
    \right]    \label{eq_dynamic_matrix_momentum}
\end{align}
Eq. \eqref{eq_dynamic_matrix_momentum} is the \textit{generic form of the dynamical matrix in momentum space under SHA}. $U(\mathbf{q})$ is the Fourier transform of the system's potential energy:
\begin{equation}
    U(\mathbf{q}) = \sum_{\mathbf{R}} U(\mathbf{R})
    e^{-i \mathbf{q} \cdot \mathbf{R}}
\end{equation}

\subsection{2D elastic theory and elastic constant for the triangular lattice}
 Strain tensor defined as 
 \begin{equation}
    \epsilon_{\alpha \beta} (\mathbf{r}) = \frac{1}{2} 
    \left( \frac{\partial u^\alpha(\mathbf{r})}{\partial r^\beta} + \frac{\partial u^\beta(\mathbf{r})}{\partial r^\alpha} 
    \right), \ \ \alpha,\beta=x,y
    \label{eq_stain_tensor}
\end{equation}
 measures the local deformation of the lattice around $\mathbf{r}$.
 $\epsilon_{xx}$ and $\epsilon_{yy}$ are the normal strain (force acted on the considered surface is perpendicular), while $\epsilon_{xy}=\epsilon_{yx}$ are the shear strain (force acted on the surface is parallel). There are two elastic constants, $\mu$ (shear modulus) and $\lambda$ (Lamé constants), of a 2D lattice. $\lambda$ is not a direct physical observable, but $B = \mu + \lambda$ is the bulk modulus when the lattice is pressed or stretched. For a 2D isotropic crystal, the deformation energy density in terms of its stain tensor $\epsilon$ and Lamé constants is:
\begin{align}
    f &= \frac{1}{2} \lambda  (\epsilon_{\alpha\alpha})^2 + \mu\ \epsilon_{\beta\gamma} \epsilon_{\beta\gamma}\\
    &= \frac{\lambda}{2} (\epsilon_{xx} + \epsilon_{yy})^2 + \mu \left( \epsilon_{xx}^2 + \epsilon_{yy}^2 + 2\epsilon_{xy}^2 \right)
    \label{eq_deformE_lame}
\end{align}

The same energy density in Voigt notation is:
\begin{align}
    f &= \frac{1}{2}
\begin{bmatrix}
\epsilon_{xx} & \epsilon_{yy} & \gamma_{xy}
\end{bmatrix}
\begin{bmatrix}
c_{11} & c_{12} & 0 \\
c_{12} & c_{11} & 0 \\
0 & 0 & c_{66}
\end{bmatrix}
\begin{bmatrix}
\epsilon_{xx} \\
\epsilon_{yy} \\
\gamma_{xy}
\end{bmatrix}, \ \gamma_{xy} = 2\epsilon_{xy} \\
\label{eq_deformE_voigt}
&= \frac{1}{2} c_{11}\ (\epsilon_{xx}^2+\epsilon_{yy}^2) + c_{12}\ \epsilon_{xx}\epsilon_{yy} + 
    2c_{66}\ \epsilon_{xy}^2
\end{align}
by comparing Eq. \eqref{eq_deformE_lame} and \eqref{eq_deformE_voigt},
\begin{equation}
   \left\{
   \begin{array}{c}
   \lambda + 2\mu = c_{11} \\
   \lambda = c_{12} \\
   \mu = c_{66}
   \end{array}
   \right.\hspace{10pt}
   \left\{
   \begin{array}{c}
   \mu = c_{66}\\
   \lambda = c_{11} - 2\ c_{66} 
   \end{array}
   \right.
   \label{eq_lame_and_voigt}
\end{equation}

The dynamical matrix Eq. \eqref{eq_dynamic_matrix_momentum} in the long-wavelength limit (in unit of energy/Area) with only quadratic-order terms can be expressed by Voigt notation:
\begin{equation}
    D_{\alpha\beta} (\mathbf{q}) = (c_{11} - c_{66}) q_\alpha q_\beta + c_{66} q^2 \delta_{\alpha \beta}
\end{equation}
\begin{equation}
    c_{11} = \frac{D_{xx} (q_x \hat{\mathbf{x}})}{q_x^2}, \ \ c_{66} = \frac{D_{xx} (q_y \hat{\mathbf{x}})}{q_y^2 }
    \label{eq_elastic_moduli}
\end{equation}
whose effective scale is $q \ll Q$, i.e. $q\ll 1/a = 1/\sqrt{A_c}$. 

In the harmonic expansion, the cost for potential energy (deformation energy) of small displacements from equilibrium positions is:
\begin{align}
    \Delta E &= \frac{1}{2} \sum_{\mathbf{q}} u_\alpha(-\mathbf{q}) D_{\alpha\beta}(\mathbf{q}) u_\beta(\mathbf{q}) \\
    D_{\alpha \beta}(\mathbf{q}) 
    &= \frac{1}{A} \sum_{\mathbf{Q}} 
    \left[
    (q_\alpha+Q_\alpha)(q_\beta+Q_\beta) U(\mathbf{q} + \mathbf{Q})
    - Q_\alpha Q_\beta U(\mathbf{Q})
    \right] \label{eq_dynamical_matrix}
\end{align}
$U(\mathbf{q})$ is the total potential energy in momentum space, which is a radial function in an isotropic system as $U(\mathbf{Q}) = U(Q)$ with $Q=|\mathbf{Q}|$.
\begin{equation}
    U(\mathbf{q} + \mathbf{Q}) = U(Q) + \left. q_\alpha \frac{\partial U(q')}{\partial q'_\alpha} \right|_{q'=Q}
+ \frac{1}{2}  q_\alpha q_\beta \left. \frac{\partial^2 U(q')}{\partial q'_\alpha \partial q'_\beta} \right|_{q'=Q} + \cdots + \frac{1}{n!}  (q_{\alpha_1} q_{\alpha_2}...q_{\alpha_n}) \left. \frac{\partial^n U(q')}{\partial q'_{\alpha_1} \partial q'_{\alpha_2}...\partial q'_{\alpha_n}} \right|_{q'=Q}
\end{equation}

$0_{th}$ order of $D^{(0)}_{\alpha\beta}$ cancels.
\begin{align}
    D^{(1)} _{\alpha \beta}(\mathbf{q}) &= \frac{1}{A} \sum_{\mathbf{Q}} \left[ (q_\alpha Q_\alpha + q_\beta Q_\beta) U(Q) + Q_\alpha Q_\beta \frac{U'(Q)}{Q}\ \mathbf{q}\cdot \mathbf{Q} \right] = 0\\
    D^{(2)} _{\alpha \beta}(\mathbf{q}) &= \frac{1}{A} \sum_{\mathbf{Q}} \left[ q_\alpha q_\beta U(Q) + (q_\alpha Q_\alpha + q_\beta Q_\beta) q_\gamma Q_\gamma
    \frac{U'(Q)}{Q} + \frac{1}{2} Q_\alpha Q_\beta q_\mu q_\nu \frac{\partial^2 U(Q)}{\partial Q_\mu \partial Q_\nu}
    \right]
\end{align}
The first order $D^{(1)} _{\alpha \beta}(\mathbf{q})$ is odd-rank related to $\mathbf{Q}$ and vanishes because of the inversion symmetry of a Bravais lattice. Hence the first non-vanishing term is the second order:
\begin{align}
    D_{\alpha \beta}(\mathbf{q}) \approx D^{(2)} _{\alpha \beta}(\mathbf{q}) = &\frac{1}{A} \sum_{\mathbf{Q}} \Bigg\{ q_\alpha q_\beta U(Q)+ \left( q_\alpha Q_\alpha + q_\beta Q_\beta \right) \left( q_x Q_x + q_y Q_y \right) \frac{U'(Q)}{Q} \notag\\ &+ \frac{1}{2} Q_\alpha Q_\beta (q_x^2 +  q_y^2)\frac{U'(Q)}{Q} + \frac{1}{2} Q_\alpha Q_\beta \left( q_x Q_x + q_y Q_y \right)^2 \frac{QU''(Q)-U'(Q)}{Q^3} \Bigg\}
\end{align}

The two elastic moduli can both be computed from $D_{xx}$:
\begin{align}
    D_{xx}^{(2)}(\mathbf{q}) = \frac{1}{A} \sum_{\mathbf{Q}} &\Bigg\{ q_x^2 U(Q) + 2q_x Q_x \left( q_x Q_x + q_y Q_y \right) \frac{U'(Q)}{Q} + \frac{1}{2} Q_x^2 (q_x^2 +  q_y^2)\frac{U'(Q)}{Q}\notag \\ 
    & + \frac{1}{2} Q_x^2 \left( q_x Q_x + q_y Q_y \right)^2 \frac{QU''(Q)-U'(Q)}{Q^3} \Bigg\}
\end{align}
According to Eq. \eqref{eq_elastic_moduli},
\begin{align}
    c_{11} &= \frac{\nu^*}{2\pi l_B^2 M} \sum_{\mathbf{Q}} \Bigg\{ U(Q) + \frac{5}{2} Q_x^2 \frac{U'(Q)}{Q} + \frac{1}{2} Q_x^4\ \frac{QU''(Q)-U'(Q)}{Q^3} \Bigg\}\\
    c_{66} &= \frac{\nu^*}{4\pi l_B^2 M} \sum_{\mathbf{Q}} \Bigg\{ Q_x^2 \frac{U'(Q)}{Q} + Q_x^2 Q_y^2  \frac{QU''(Q)-U'(Q)}{Q^3} \Bigg\}
\end{align}
Finally the analytical expression of the elastic constants up to the second order can be computed as:
\begin{align}
    \mu &= c_{66} = \frac{\nu^*}{4\pi l_B^2 M} \sum_{\mathbf{Q}\ne 0} Q_x^2\Bigg\{ \frac{U'(Q)}{Q} + Q_y^2  \frac{QU''(Q)-U'(Q)}{Q^3} \Bigg\}\\
    \lambda &= c_{11} - 2c_{66} = \frac{\nu^*}{2\pi l_B^2 M} \sum_{\mathbf{Q}\ne 0} \Bigg\{ U(Q) + \frac{3}{2} Q_x^2 \frac{U'(Q)}{Q} +  Q_x^2 (\frac{Q_x^2}{2}-Q_y^2) \frac{QU''(Q)-U'(Q)}{Q^3}\Bigg\}
\end{align}
Note: here the elastic constants is defined for per unit cell.

\section{Defect-mediated melting}
A dislocation is a distortion of a perfect crystal. The Burgers vector $\vec{b}$ measures the lattice distortion caused by the presence of defects. The Berezinskii-Kosterlitz-Thouless (BKT) transition describes the melting of two-dimensional crystals (or superfluids, XY models, etc.) driven by the unbinding of topological defects. It is topological in nature, marked by defect unbinding and universal jump conditions in elastic constants. 
In a 2D solid, translational order is not destroyed by ordinary thermal fluctuations as it is in 3D, but by the proliferation of dislocations and disclinations. 
The Kosterlitz--Thouless--Halperin--Nelson--Young (KTHNY) theory provides a renormalization-group (RG) framework for this process. In an idealized clean system, the whole KTHNY melting process is two-step: topological quasi-long-range order $\rightarrow$ intermediate hexatic phase $\rightarrow$ liquid phase, which could merge into a single transition because of the disorder and other effects in a real system.

\subsection{RG flow equations}
For an isotropic 2D solid, the free‑energy cost of long‑wavelength deformations is
\begin{equation}
  F [\mathbf u]= \mu u_{ij}^{2}+ \frac{\lambda}{2}u_{kk}^{2}, \qquad
  u_{ij}=\tfrac12(\partial_i u_j+\partial_j u_i)
  \label{eq_F_el}
\end{equation}
where $\mu$ and $\lambda$ are the in‑plane shear and the second Lam\'e moduli, respectively.  The combination
\begin{equation}
  \label{eq_young}
  Y \equiv \frac{4\mu (\mu+\lambda)}{2\mu+\lambda}
\end{equation}
is the 2D Young's modulus. For a triangular lattice with lattice constant $a$, it is conventional to define a dimensionless coupling $K$, first introduced by Halperin and Nelson as the \emph{2D Cauchy modulus}, so that its (renormalized) value reaches a universal constant when melting.
\begin{equation}
    K = \frac{a^2 Y}{k_B T}
\end{equation}

We introduce the so-called dislocation fugacity $y$ that measures the Boltzmann weight to create a (bound) dislocation pair. Write the renormalization flow variable as $l$, the coupled differential equations for $K(l)$, $y(l)$, $\mu(l)$, and $\lambda(l)$ are:
\begin{align}
    \frac{d}{dl} \left( \frac{1}{K(l)} \right) = A(K)y^2,\,
    \frac{d y(l)}{dl} = \left( 2- \frac{K(l)}{8\pi} \right)y + B(K) y^2
    \label{eq_RG_Ky}
\end{align}
\begin{equation}
  K(l)=\frac{a_0^2}{k_B T}\, Y(l),
  \qquad
  Y(l)=\frac{4\mu(l)\,[\mu(l)+\lambda(l)]}{2\mu(l)+\lambda(l)},
\end{equation}
$A(K)$ and $B(K)$ are both positive and have the explicit forms for a triangular lattice:
\begin{align}
    A(K) &=  \frac{3\pi}{4} e^{K(l)/8\pi} \{ 2 I_0\left(K(l)/8\pi \right) - I_1\left(K(l)/8\pi \right) \} \\
    B(K) &= 2\pi\ e^{K(l)/16\pi}\ I_0\left(K(l)/8\pi \right)
\end{align}
$I_0$ and $I_1$ are modified Bessel functions.

\begin{figure*}
    \centering
    \includegraphics[width=0.75\linewidth]{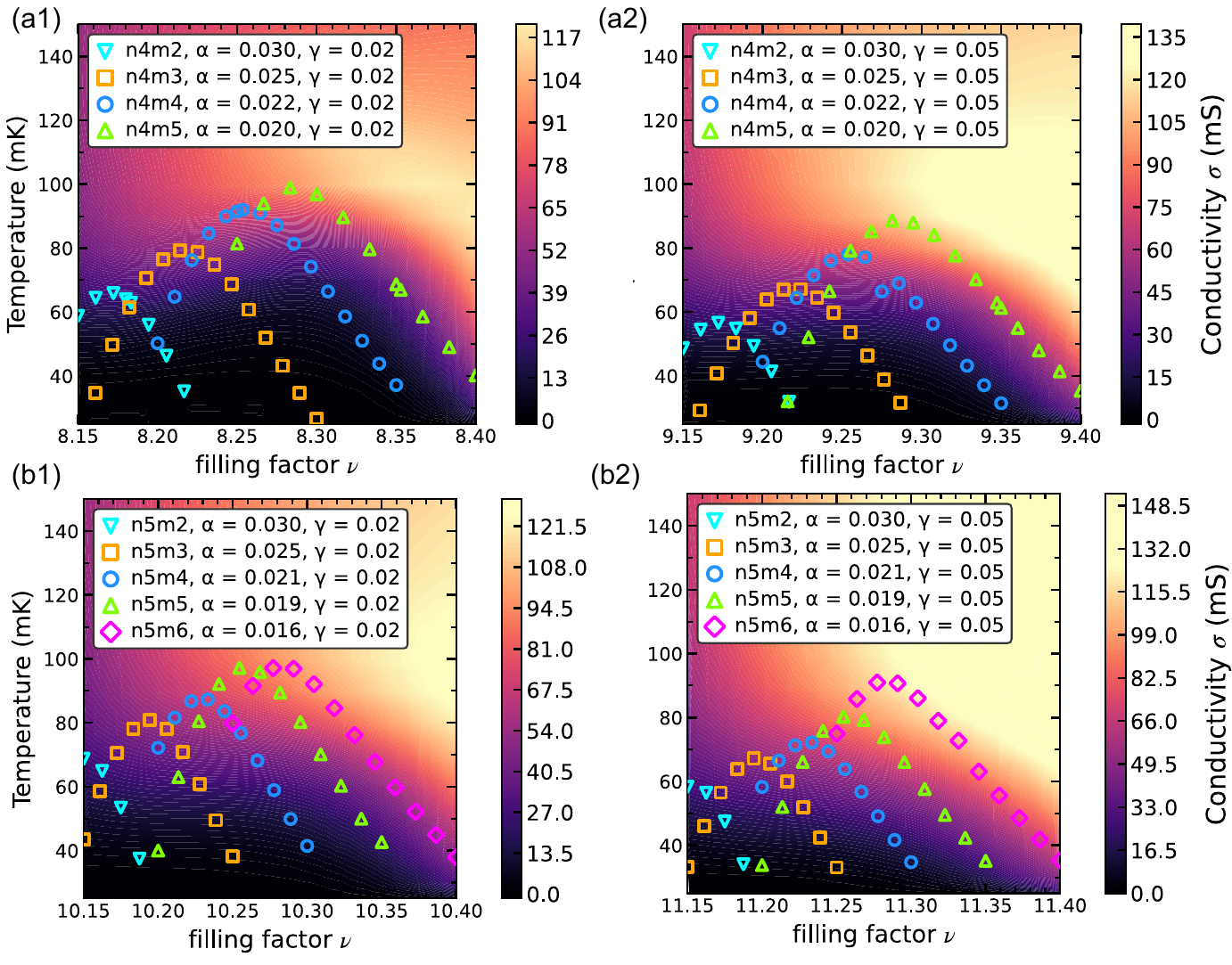}
    \caption{Color maps of the longitudinal conductance as a function of temperature $T$ and filling factor $\nu$, showing re-entrant insulating regions associated with bubble-crystal phases at LL $N=4$ and $5$. Overlaid symbols mark the melting temperatures $T_m(\nu)$ for the indicated $M$-electron bubble phases, obtained from the KTHNY--RG melting criterion using Landau-level--projected HF elastic moduli (model parameters ). $\alpha$ and $\gamma$ are listed in the legends. }
    \label{fig_s_tm}
\end{figure*}

\end{document}